\documentclass[aps, superscriptaddress ,twocolumn,prl]{revtex4}
\usepackage{amsmath}
\usepackage{amssymb}
\usepackage{epsfig}
\usepackage{graphics}
\usepackage{bm}
\usepackage{braket}

\begin{document}
\title{Harmonically trapped dipolar fermions in a two-dimensional square lattice}
\author{Anne-Louise\ Gadsb\o lle}
\affiliation{Lundbeck Foundation Theoretical Center for Quantum System Research}
\affiliation{Department of Physics and Astronomy, University of Aarhus, Ny Munkegade, DK-8000 Aarhus C, Denmark}
\author{G.\ M.\ Bruun}
\affiliation{Department of Physics and Astronomy, University of Aarhus, Ny Munkegade, DK-8000 Aarhus C, Denmark}

\begin{abstract}
We consider dipolar fermions in a two-dimensional square lattice and a harmonic trapping potential. The anisotropy of the dipolar interaction combined 
with the lattice  leads to transitions between  phases with 
density order of different symmetries. We show that the attractive part of the dipolar interaction results in a 
superfluid phase which is suppressed by density order. 
 The  trapping potential is demonstrated to make the different phases  co-exist,   forming  
ring and island structures. The phases with density and superfluid order  can overlap forming regions with  supersolid order.

\end{abstract}
\maketitle

The trapping of dipolar atoms and molecules  is a promising new 
research field. The anisotropy of the dipole interaction offers unique opportunities for exploring novel few-body~\cite{Cremon,Volosniev} and 
many-body quantum systems~\cite{BaranovRev,LahayeRev}. Experimentally, one has realized 
Bose-Einstein condensates of  $^{52}$Cr atoms~\cite{Lahaye,Koch} and 
$^{164}$Dy atoms~\cite{Lu} with large magnetic dipole moments,  as well as gases close to quantum degeneracy 
of $^{40}$K$^{87}$Rb  molecules with an electric dipole moment~\cite{KRb}. Furthermore, 
the first experimental steps toward realizing dipolar molecules in an optical lattice have recently been reported~\cite{Danzl,Jin}.
The lattice makes the physics very rich: Density ordered phases with a complicated unit cell~\cite{Mikelsons}, liquid crystal phases~\cite{Lin}, and a supersolid 
phase~\cite{He} have been predicted to exist for  fermionic dipoles in a 2D lattice with dipole moments perpendicular to the lattice plane, and tilting the 
dipoles towards the lattice plane leads to  bond-solid order and $p$-wave superfluidity at half-filling~\cite{Bhongale}.

We consider fermionic dipoles in a 2D square  lattice at zero temperature. A harmonic potential, which is always present in trapped atomic/molecular systems, 
is included exactly, since the characteristic 
lengths of the ordered phases  can be comparable to the system size for experimentally realistic systems. This 
 means that one cannot simply resort to the local density approximation.  
A main purpose of the present paper is to study the rich physics coming from the 
interplay between the anisotropic dipole interaction, the optical lattice, and the inhomogeneity induced by the trapping potential. 
For experimentally realistic systems, 
we demonstrate the existence  of  competing phases with density order of checker-board or stripe symmetry, and 
superfluid order. Due to the trapping potential, these phases can co-exist and  sometimes even spatially overlap leading to regions with supersolid order.

\begin{figure}[bh]
\includegraphics[width=1\columnwidth]{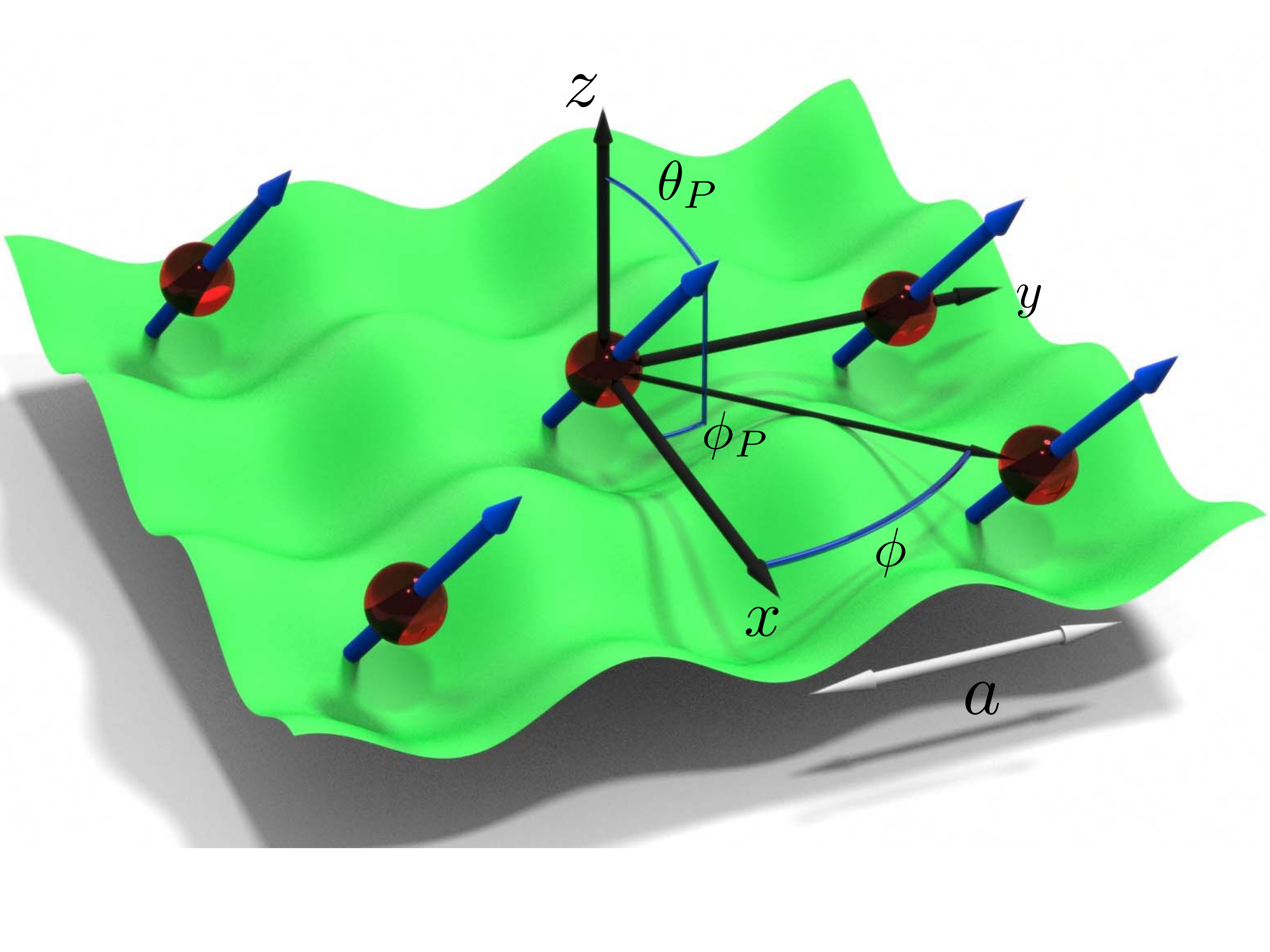}
\caption{(color on-line) We consider dipoles in a square 2D lattice in the $xy$-plane. The dipoles are aligned forming an angle $\theta_P$ with the $z$-axis and the azimuthal 
angle $\phi_P$ with the $x$-axis which is parallel to one of the lattice vectors. }
\label{setup}
\end{figure}

\noindent\emph{Basic formalism\,\,\,}
The fermions have a  mass $m$ and a dipole moment ${\mathbf d}$ 
which is aligned by an external field forming the angle $\theta_P$ with the $z$-axis perpendicular to the lattice ($xy$) plane and the 
azimuthal angle $\phi_P$ with respect to one of the lattice vectors parallel to the $x$-axis,   see Fig.\ \ref{setup}. In the lowest band approximation, 
this system is described by the extended Hubbard model with the Hamiltonian $\hat{H}=\hat{H}_0+\hat{V}$ where 
 \begin{equation}
\hat{H}_0=-t\sum_{\langle ij\rangle}\left(\hat{c}_{i}^{\dagger}\hat{c}_{j }+h.c.\right)
+\sum_{i}\left(\frac 1 2 m\omega^2r_i^2-\mu\right)\hat{n}_{i}
\end{equation}
and 
 \begin{equation}
\hat{V}=\frac 1 2 \sum_{i\neq j }V_D({\mathbf r}_{ij})\hat{n}_{i}\hat{n}_{j}
\label{Hint}
\end{equation}
with ${\mathbf r}_{ij}={\mathbf r}_{i}-{\mathbf r}_{j}$.
Here, $\hat{c}_{i}$ removes a dipole at site $i$ with position ${\mathbf r}_i$, $\hat{n}_{i}=\hat{c}_{i}^{\dagger}\hat{c}_{i}$, 
$\mu$ is the chemical potential, $t$ is the 
hopping matrix element between nearest neighbor sites $\langle ij\rangle$, and $i \neq j$ since we are considering identical fermions. The trapping frequency is $\omega$ and the 
interaction between two dipoles separated by ${\mathbf{r}}$ is  
\begin{gather}
V_D({\mathbf{r}})=\frac{D^2}{r^3}\left[1-3\cos^{2}(\theta_{\rm rd})\right]\nonumber\\
=\frac{D^2}{r^3}\left[1-3\cos^{2}(\phi_{P}-\phi)\sin^{2}(\theta_{P})\right] 
\label{Interaction}
\end{gather}
with $D^2=d^2/4\pi\epsilon_0$ for electric dipoles and $\theta_{\rm rd}$ the angle between ${\mathbf d}$ and
${\mathbf r}=r(\cos \phi, \sin \phi,0)$, see Fig.\ \ref{setup}.

Since the dipolar interaction (\ref{Interaction})  has both attractive and repulsive 
regions, the system exhibits both pairing and density instabilities depending on $(\theta_P,\phi_P)$. 
To model this complex behavior, we decouple the interaction $\hat{V}$ using the mean-field approximation 
which yields 
\begin{gather}
\hat{V}_{\rm MF}=
\sum_{i\neq j}V_D({\mathbf r}_{ij})\left( \langle \hat{n}_{j}\rangle \hat{n}_{i}-\frac{1}{2}\langle \hat{n}_{j}\rangle\langle
\hat{n}_{i}\rangle\right) \nonumber\\
+\sum_{i\neq j}\frac{V_D({\mathbf r}_{ij})}{2}
\left( \braket{\hat{c}_{j}\hat{c}_{i}}\hat{c}_{i}^{\dagger}\hat{c}_{j}^{\dagger}+h.c.-|\braket{\hat{c}_{j}\hat{c}_{i}}|^{2} \right)
\label{VMF}
\end{gather}
where $\langle\hat{c}_{j}\hat{c}_{i}\rangle$  is the pairing order parameter.
Even though fluctuations are important in 2D, we expect mean-field theory  to capture the existence 
and competition between different ordered phases at  $T=0$. Indeed, mean-field theory is widely used in the high-$T_c$
community to describe the competition between e.g.\ anti-ferromagnetic and superfluid ordering~\cite{HighTcs}.
We have not included the  Fock term in (\ref{VMF}), since one dipole interacts with many others making the problem similar 
to a high dimensional one, for   which  the Hartree term dominates~\cite{Muller}. The 
Fock term  has recently been shown to lead to  bond-solid order phases at half-filling~\cite{Bhongale}.

The mean-field Hamiltonian $\hat H_0+\hat{V}_{\rm MF}$ is diagonalized by the Bogoliubov transformation 
$\hat{c}_{ i}= \sum_{E_\eta>0} (u_{\eta }^{i}\hat{\gamma}_{\eta }+v_{\eta }^{i *} \hat{\gamma}_{\eta }^{\dagger})$
where $\gamma_{\eta}$ are fermionic  operators annihilating a quasi-particle with energy $E_\eta$. The 
 wave functions $u_{\eta }^{i}$ and $v_{\eta }^{i }$ satisfy the 
Bogoliubov-de Gennes equations
\begin{align}
\sum_j\begin{pmatrix}
L_{ij} & \Delta_{ij} \\
\Delta_{ji}^* & -L_{ij}
\end{pmatrix}
\begin{pmatrix}
u_{ \eta }^{j} \\
v_{ \eta}^{j}
\end{pmatrix}
=E_{\eta }
\begin{pmatrix}
u_{ \eta }^{i} \\
v_{\eta }^{i}
\end{pmatrix},
\end{align}
with $\Delta_{ij}=V_D({\mathbf r}_{ij})\langle \hat c_{j}\hat c_{i}\rangle$ and 
\begin{gather}
L_{ij}=-t \delta_{\langle ij\rangle}+(\sum_{k}V_D({\mathbf r}_{ik})\langle n_{k}\rangle+\frac m 2 \omega^2r_i^2-\mu) \delta_{ij}.
\end{gather}
Here $\delta_{ij}$ and $\delta_{\langle ij\rangle}$ are
the Kronecker delta functions connecting on-site and nearest neighbor sites, respectively.
Self-consistency is obtained iteratively through the usual relations: 
$\langle\hat{n}_{i}\rangle=\sum_{E_\eta>0}|v_{\eta}^{i}|^{2}$ and
 $\langle\hat{c}_{i }\hat{c}_{j }\rangle =\sum_{E_\eta>0}u_{\eta}^{i}v_{\eta}^{j*} $.

\begin{figure}[th]
\includegraphics[clip=true,width=0.85\columnwidth]{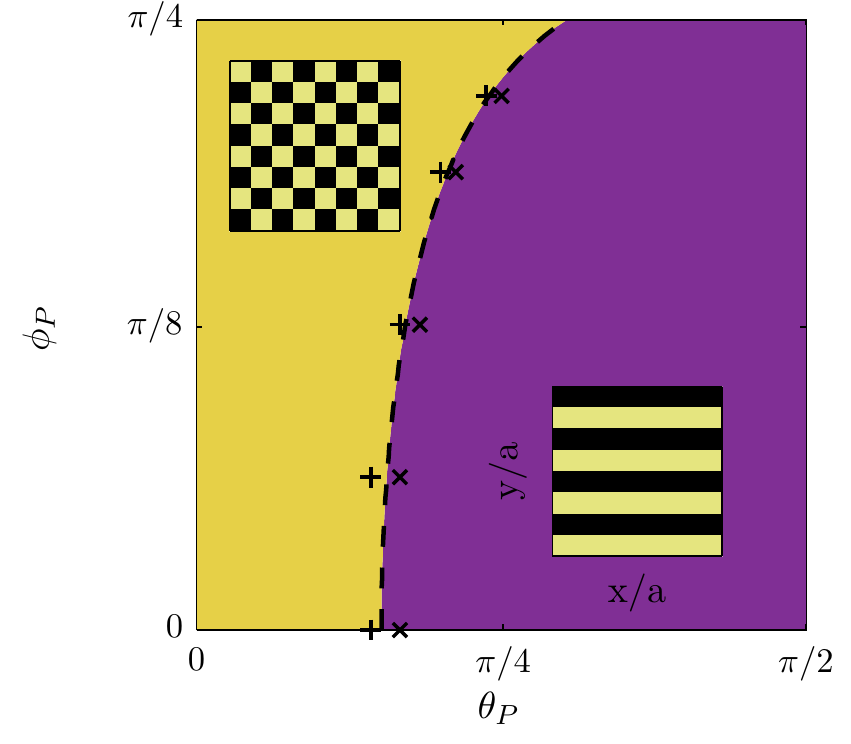}
\includegraphics[clip=true,width=0.8\columnwidth]{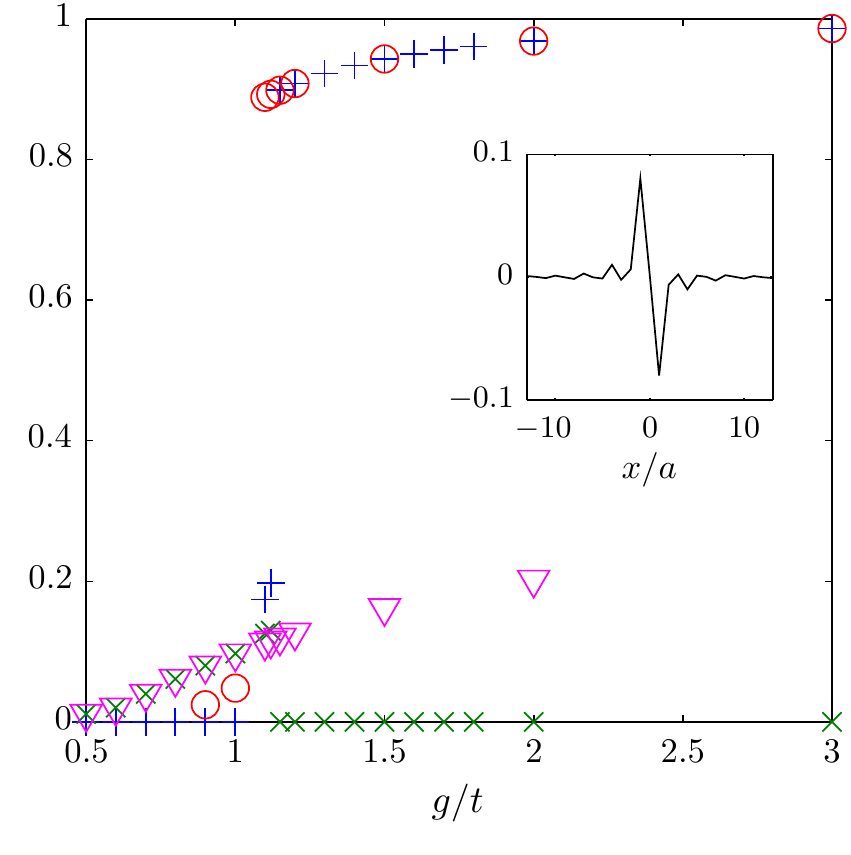}
\caption{(color on-line) Top: The phase-diagram for an untrapped system for strong coupling $g/t=16$ and half-filling.
 The dashed line shows the boundary between the checker-board and the striped phase.  Bottom: The  nearest neighbor pairing 
$|\langle \hat c_{i+\hat x}\hat c_{i}\rangle|$ ($\times$'s) and stripe order $|\langle\hat n_{i}-\hat n_{i+\hat y}\rangle|$ ($+$'s) as a function of $g/t$
 with $(\theta_{P},\phi_P)=(\pi/2,0)$. 
The $\circ$'s show $|\langle n_{i}-n_{{i}+\hat y}\rangle|$ when we take the gap to be zero. The $\triangledown$'s show $|\langle \hat c_{i+\hat x}\hat c_{i}\rangle|$ when the density is taken 
to be homogeneous. The inset shows a cut of $\langle \hat c_{i+\hat x}\hat c_{i}\rangle$ along the $x$-axis when $g/t=0.9$. }
\label{phasediagram}
\end{figure}

\noindent\emph{No trapping potential\,\,\,}
Consider first a system with no trapping potential. We plot in Fig.\ \ref{phasediagram} the 
ground state  as a function of the dipolar angles $(\theta_P,\phi_P)$ in the limit of strong interactions $g/t=16$ with 
$g=D^2/a^3$ where $a$ is the lattice constant. The filling fraction is $f=N^{-1}\sum_i\langle\hat n_i\rangle=1/2$ with $N$ the number of lattice sites. 
We  consider two possible ground states: One with a checker-board density order and one with a striped density order. 
The $+$ symbols and the $\times$ symbols indicate when mean-field theory on a $26 \times 26$ lattice
predicts the  checker-board phase and the striped phase  to have the lowest energy, respectively. In order to take into account the long-range nature of the dipolar interaction we have calculated the Hartree term by duplicating the $26 \times 26 $ lattice so that it constitutes a lattice of $9 \cdot 26 \times 9 \cdot  26$ sites. 
For small $\theta_P$, the checker-board phase is favored whereas for larger 
$\theta_P$ a phase with stripes along the $x$-direction has the lowest energy. This is easily understood for  $\phi_P=0$, 
where the dipoles are aligned head-to-tail in this phase 
which clearly minimizes the interaction energy. When $\phi_P>0$, the alignment is not perfect by stripe formation along the $x$-axis, but the interaction 
energy is still minimized although it requires larger tilting angles $\theta_P$ as expected. This is illustrated further by the dashed line
  separating the checker-board phase from the striped phase, which  is a result of comparing the classical 
 interaction energy in the two phases.  The good  agreement between the numerics and this calculation 
demonstrates that for strong coupling $g/t\gg 1$ the kinetic energy can be neglected,
and the problem becomes classical.

When $\theta_P\ge \arcsin(1/\sqrt 3)\approx 0.2 \pi$, the interaction has attractive regions, and we 
 now examine the competition between the resulting pairing instability and the  instability towards density order. 
In Fig.\ \ref{phasediagram}, we plot as $\times$'s  the largest value of the nearest neighbor 
pairing $|\langle \hat c_{i+\hat x}\hat c_{i}\rangle|$ as a function of the coupling strength $g/t$
with $\hat x$ a unit vector along the $x$-direction. We have chosen 
  $(\theta_P,\phi_P)=(\pi/2,0)$ and   $f=1/3$. The calculations are performed 
on a $27\times 27$ lattice. For weak coupling, the ground state is  a superfluid. The cross section 
along the $x$-axis of the pair wave function $\braket{\hat{c}_{j}\hat{c}_{i}}$ with ${\mathbf{r}}_{i}=(0,0)$  plotted in the inset,
illustrates that it is odd under inversion as for the homogeneous case~\cite{BruunTaylor,Baranov}. For simplicity,
we refer to this as $p$-wave symmetry in the following, even though the pair wave function in general contains higher odd
components of angular momenta. 
 For stronger coupling, the pairing vanishes as the striped phase emerges. The stripe order 
 defined as $\langle\hat n_{i}-\hat n_{i+\hat y}\rangle$, with $\hat y$ a 
unit vector along the y-direction so that $\langle\hat n_{i}\rangle$ and $\langle\hat n_{i+\hat y}\rangle$
give the densities for a site in the stripe and  next to the stripe respectively,
 is plotted as $+$'s in Fig.\ \ref{phasediagram}.
We also plot as $\circ$'s the stripe order parameter when pairing is not included in the calculation, and as $\triangledown$'s the pairing order parameter when the stripe formation is 
not included. This demonstrate that the stripe  order is insensitive to pairing since the critical coupling strength for the formation of stripes 
essentially does not 
change when pairing is included. The pairing on the other hand does not vanish with increasing coupling if the stripes are suppressed. We therefore conclude that  
it is the pairing which is suppressed by the stripe formation and not the other way around.

More complicated  density order with larger unit cells can appear for untrapped systems, as has recently been demonstrated for $\theta_P=0$~\cite{Mikelsons}.
Here, we  focus on the  experimentally most relevant  orders  with the smallest unit cell, since the trapping potential will complicate
the observation of orders with a large unit cells.

\noindent\emph{Trapped case\,\,\,}
We now examine the interplay between the harmonic trapping potential and the competition between density and pairing instabilities.

\begin{figure}
\includegraphics[clip=true,width=1\columnwidth]{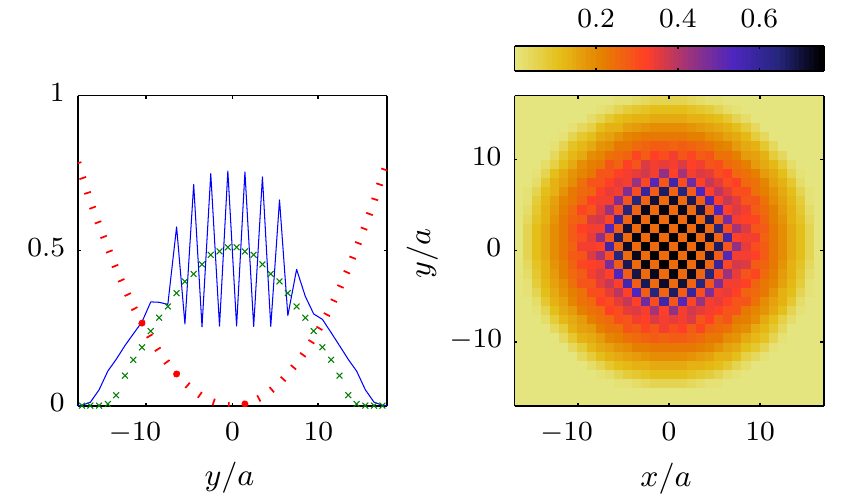}
\caption{(color on-line)  Left: The density profile through the center of the trap. 
The blue solid line is for $g/t=1$ and $207$ particles trapped and the green crosses are for $g/t=0.5$ and $144$ particles trapped.
 The red dotted line indicates the trapping potential in units of 2.5t with $\tilde{\omega}=0.24$. Right: 2D plot of the density in the lattice plane for the 
 $g/t=1$ case.  }
\label{Traptheta0}
\end{figure}

\emph{Angle $\theta_P=0$\,\,\,} 
Consider first the case when the dipolar orientation is perpendicular to the lattice so that the interaction is purely repulsive.  Figure \ref{Traptheta0} shows the density profile for trapped dipoles with $\tilde\omega=\omega a\sqrt{m/t}= 0.24$ which is a relatively weak trapping
 potential  so that there are  regions with phases resembling those for the case with 
 no trapping potential. 
 For $g/t=0.5$ and $144$ particles, there is no density order whereas for stronger 
coupling $g/t=1$ and $207$ particles, the system exhibits a checker-board density profile in the center of the trap. In this region, $f\simeq1/2$ which
 is optimal for the checker-board phase~\cite{Mikelsons}. The checker-board phase is surrounded by a normal phase with a lower density. 
Experimentally, similar shell structures have been observed   for atoms with a short range interaction~\cite{BlochRMP}. 

\begin{figure}
\includegraphics[clip=true,width=1\columnwidth]{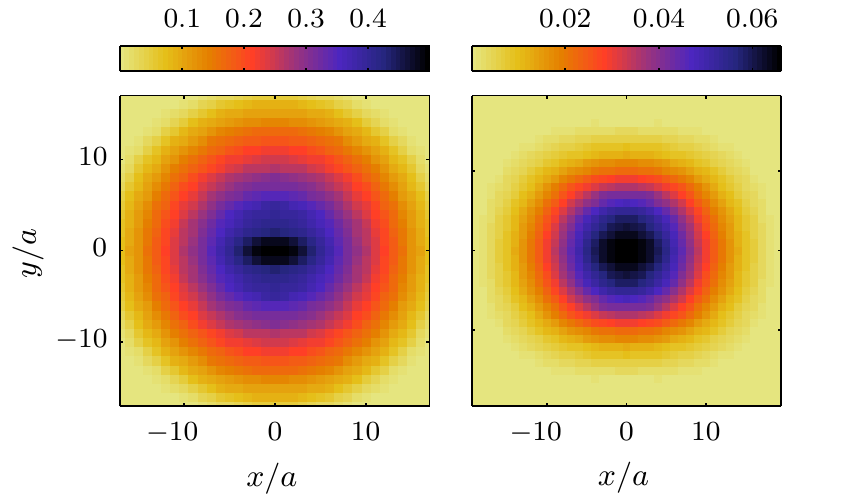}
\caption{(color on-line) Left: The particle density for 205 dipoles  with $\theta_P=\pi/2$, $g/t=0.85$, and $\tilde\omega=0.11$. Right: The nearest neighbor pairing $(|\langle \hat c_{i+\hat x}\hat c_{i}\rangle|+|\langle \hat c_{i-\hat x}\hat c_{i}\rangle|)/2$.
}
\label{Trapthetapi/2Weak}
\end{figure}

\emph{Angles $(\theta_P,\phi_P)=(\pi/2,0)$\,\,\,}
Consider next the  case when the dipoles are aligned in the plane along the $x$-axis ($\phi_P=0$).
 There is then a 
competition between density and pairing order. In Fig.\ \ref{Trapthetapi/2Weak}, we plot the density and the nearest neighbor pairing 
$(|\langle \hat c_{i+\hat x}\hat c_{i}\rangle|+|\langle \hat c_{i-\hat x}\hat c_{i}\rangle|)/2$ (symmetrized to reduce trap effects)
 for $g/t=0.85$, $\tilde w=0.11$ and $205$ dipoles trapped in a  
 $39 \times 39$ lattice.
For this set of parameters, 
there is no density order since the density is low and the interaction weak. This results in $p$-wave
 pairing throughout most of the cloud. The cloud profile is slightly elongated in the $x$-direction due to the anisotropy of the 
 dipolar interaction in analogy with what has been observed for dipolar condensates~\cite{LahayeRev}. 

\begin{figure}
\includegraphics[clip=true,width=1\columnwidth]{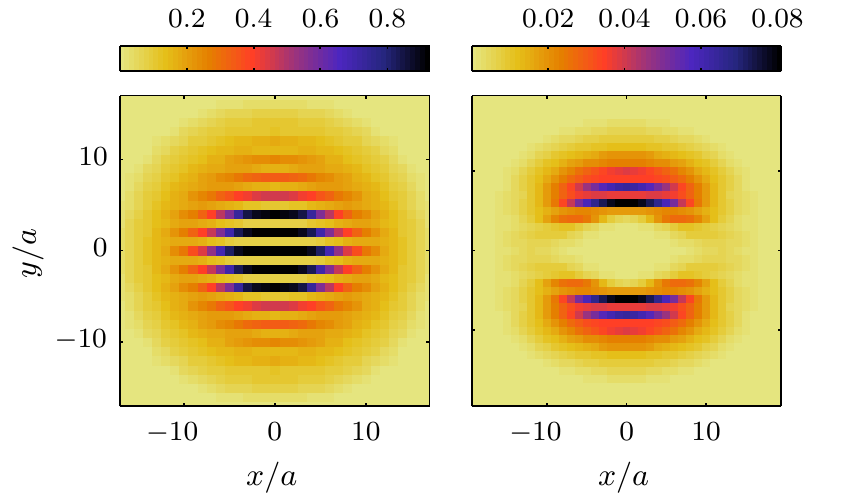}
\caption{(color on-line) Left: The density profile for 180 dipoles  with $\theta_P=\pi/2$, $g/t=1$, and $\tilde\omega=0.11$. Right:  The nearest neighbor pairing $(|\langle \hat c_{i+\hat x}\hat c_{i}\rangle|+|\langle \hat c_{i-\hat x}\hat c_{i}\rangle|)/2$.}
\label{Trapthetapi/2Strong}
\end{figure}

Figure \ref{Trapthetapi/2Strong} depicts the density and the nearest neighbor pairing for a stronger coupling strength $g/t=1$ with  $\tilde w=0.11$ and $180$ dipoles trapped on a $39 \times 39$  lattice. 
There is a pronounced stripe  order in the center of the trap with an average filling fraction $f\simeq 1/2$
which  squeezes the pairing away from the center into two islands centered at $x=0$. 
This intriguing island structure is a consequence of the anisotropy of the interaction. Since the interaction 
is attractive when the dipoles are aligned head-to-tail ($x$-direction) and repulsive when they are side-by-side ($y$-direction), 
the average interaction for a given radius is attractive in the regions close to the $y$-axis whereas it is repulsive 
 in the regions close to the $x$-axis. Thus, pairing can exist in the islands around $x=0$ away from the center where stripe order 
 dominates, whereas it is suppressed in the regions around $y=0$.
 These  islands  of pairing should be compared with the ring structures  predicted for a two-component trapped
fermi gas with a short range isotropic repulsive interaction where anti-ferromagnetic order competes with $d$-wave pairing~\cite{AndersenBruun}. 
Remarkably, the stripe order is not completely suppressed in the two islands of pairing.
The pairing in fact  oscillates \emph{in-phase} with the stripe order. This indicates that
the trapping potential induces regions of pairing co-existing with density order as in a supersolid -- 
a phase which has been subject to intense investigations since its theoretical prediction long ago~\cite{Andreev,Leggett,Chan,Capogrosso,DeMelo}. 

The lattices considered here have experimentally realistic  sizes, and with  
 electric dipole moments up to several Debye for the experimentally relevant  KRb, RbCs, and  LiCs molecules~\cite{KRb,Pilch,Kotochigova,Deiglmayr},
 one can easily reach the strong-coupling regime with $g/t \gg 1$ using typical values for an 
optical lattice. The density ordering can be observed directly by in-situ imaging~\cite{Sherson}
or by time-of-flight experiments which also can detect  pairing correlations~\cite{Altman,AndersenBruun}.

In conclusion, we have shown that dipolar fermions at $T=0$ in a 2D square lattice exhibit  phases of density order
with different symmetries as well as a $p$-wave superfluid phase. 
The system is unstable towards pairing when the interaction has attractive channels. However, any superfluidity is suppressed if there is an instability towards stripe formation such as in regions 
with a density close to half filling. The trapping potential leads to an inhomogeneous filling fraction which results in the co-existence of several phases. 
These phases can overlap resulting in the presence of regions with supersolid order.

We are grateful to N.\ Zinner, M.\ Valiente, and L.\ Mathey for useful discussions. Furthermore, A.-L. G. is grateful to N.\ Nygaard for helping setting up the initial numerics and to S.\ Gammelmark and O.\ S\o e S\o rensen  for assistance with Fig. \ref{setup}.

\end{document}